\documentclass[aps,showpacs,prb,twocolumn,floatfix]{revtex4}
\usepackage{amsmath}
\usepackage{amssymb}
\usepackage{graphicx}
\usepackage{dcolumn}
\usepackage{bm}

\begin{document}

\title{Spin transmission control in helical magnetic fields}

\author{Henri Saarikoski}
\email[Electronic address: \\]{henri.saarikoski@gmail.com}
\affiliation{Department of Theoretical Physics, Regensburg University, 93040 Regensburg, Germany}
\author{Tobias Dollinger}
\affiliation{Department of Theoretical Physics, Regensburg University, 93040 Regensburg, Germany}
\author{Klaus Richter}
\affiliation{Department of Theoretical Physics, Regensburg University, 93040 Regensburg, Germany}

\begin{abstract}
We calculate spin transport in two-dimensional waveguides in the
presence of spatially modulated Zeeman-split energy bands.
We show that in a regime where the spin evolution is predominantly adiabatic
the spin backscattering rate can be tuned via diabatic Landau-Zener
transitions between the spin-split bands
[C. Betthausen {\em et. al.}, Science {\bf 337}, 324 (2012)].
This mechanism is tolerant against spin-independent scattering processes.
Completely spin-polarized systems show full spin backscattering, and thus current switching.
In partially spin-polarized systems a spatial sequence of Landau-Zener transition points
enhances the resistance modulation via reoccupation of backscattered
spin-polarized transport modes.
We discuss a possible application as a spin transistor.
\end{abstract}

\pacs{71.70.Ej, 72.25.Dc, 85.75.Hh}

%72.25.Dc - Spin polarized transport in semiconductors
%85.75.Hh - Spin polarized field effect transistors
%71.70.Ej   - Spin-orbit coupling, Zeeman and Stark splitting, Jahn-Teller effect. 

\maketitle

\section{Introduction}
\label{sec:intro}

Current technologies in semiconductor electronics have fundamental limits on the transistor switching
times that can be achieved with low energy consumption.
Integration of electron spin-based functionalities into devices may
lead to faster operation.\cite{Awschalom2002}
%Electron spin can potentially be manipulated with far smaller energies than charge.
Datta and Das proposed an idea to modulate current in a spin
transistor device via spin precession in a spin-orbit field.\cite{Datta1990}
In their concept spin is injected from a ferromagnetic source into a channel of a two-dimensional electron gas (2DEG)
where spin precesses in a gate-controlled spin-orbit field. The drain
is another ferromagnet where spin magnetic moment orientation parallel (antiparallel) to the
magnetization direction of the drain corresponds to the transistor 'on' ('off') position.
Signatures of the Datta-Das spin transistor mechanism have been demonstrated in a non-local measurement.\cite{Koo2009}
However, signal levels are small due to issues concerning spin injection efficiency and fast spin relaxation.\cite{Fabian2004}
Fast spin decay makes information encoded in spin very volatile and limits its transmission range.

Recently an alternative way to achieve spin transistor action has been proposed: 
stability of spin is enhanced by keeping spin transport in the adiabatic regime\cite{Born1928} and spin transmission can then be
controlled via Landau-Zener transitions in spatially modulated spin-split bands.\cite{betthausen}
This leads effectively to a tunable backscattering of spins which
changes conductance and the degree of spin polarization of transmitted electrons in the device.
The validity of this approach was shown in transport experiments\cite{betthausen} in magnetically modulated diluted (Cd,Mn)Te
magnetic semiconductor quantum wells where the \textit{s-d}~exchange interaction between electronic states and the
localized magnetic moments of the Mn atoms gives rise to an enhanced {g-factor}
and a giant Zeeman splitting.\cite{Furdyna1988}
In the low-field limit at low temperatures the g-factor is approximately constant with values ranging up to
several hundreds. In these experiments spin transistor action was realized by combining helical and tunable
homogeneous magnetic field components. The helical field component was created by placing a premagnetized
ferromagnetic stripe grating above the sample surface. A dysprosium stripe grating induces a stray field which is
approximately helical in the plane of the 2DEG with a field strength of the order of 50 mT.
Due to the giant g-factor the spin polarization of the ground state in this field was about $10\%$.

Motivated by these experiments we consider here spin transmission control in helical magnetic fields
in the presence of spin-independent disorder scattering and magnetic field coupling to orbital dynamics.
We show that current switching can be attained in fully spin-polarized systems despite disorder.
This finding is in contrast to the Datta-Das spin transistor; its operation is disrupted if the mean free path of electrons
is of the order of the channel length.  We further show that transitions between transport modes induced by
orbital coupling, which was not considered in
Ref.~\onlinecite{betthausen},
may enhance the resistance modulation in partially polarized waveguides.

\section{Model}
\label{sec:model}

\subsection{Effective mass Hamiltonian}

\begin{figure*}
\includegraphics[width=2.075\columnwidth]{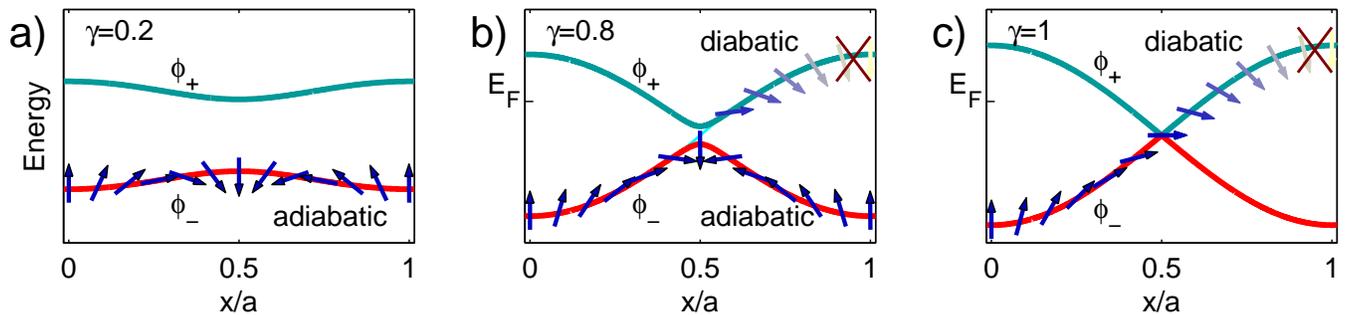}
\caption{(Color online) Adiabatic and diabatic spin evolution for a two-level system described by Eqs.\ (\ref{eq:ez}) and (\ref{eq:theta})
for $B_{\mathrm{c}}/B_{\mathrm{helix}}=\gamma=0.2$ (a), $\gamma=0.8$ (b), and $\gamma=1$ (c).
Magnetic field orientation at $x=0$ is down for $g_{\mathrm{eff}}>0$.
Fast rotation of the  $\phi_-$ eigenstate spin orientation (arrows attached to the red line) may lead
to a diabatic transition to $\phi_+$ at a level anticrossing (b).
If $E_{\mathrm{F}}$ is lower than the energy of $\phi_+$ the wave function decays and the spin is backscattered
at the corresponding potential barrier.
At $\gamma=1$ the energy levels cross at $x=a/2$ (c). Spin which is transported to this point adiabatically in the $\phi_-$ state has
a wave function overlap of 1 with $\phi_+$, and a diabatic transition occurs followed by backscattering.}
\label{fig:fig0}
\end{figure*}
We use an effective mass model to describe electrons moving in an $xy$-plane of a 2DEG in a waveguide.
The orbital motion couples to the magnetic field component perpendicular
to the plane of the 2DEG  ($z$-direction). We neglect electron-electron interactions and spin-orbit coupling.
The effective mass Hamiltonian is then
\begin{equation}
\mathrm{\hat{H}} =
\frac{1}{2 m^*}\ \left(\hat{\bf{P}} - \mathrm{e} \bf{A}(\bf{x})\right)^2 + \frac{1}{2} {g}_{\mathrm{eff}} \mathrm{\mu_{\mathrm{B}}}\mathbf{B}(\mathbf{x})\cdot \bm{\sigma}+V_\mathrm{dis}({\mathbf x}),
\label{eq:h}
\end{equation}
where $\hat{\bf{P}}$ is the momentum operator, $m^*$ the effective mass, $\mathbf{B}(\mathbf{x})$ the magnetic field,
$\mathbf{A}(\mathbf{x})$ the vector potential of the $z$-component of $\mathbf{B}(\mathbf{x})$, 
${g}_{\mathrm{eff}}$ the effective g-factor,
$\mathrm{\mu_{\mathrm{B}}}$ the Bohr magneton,
$\bm{\sigma}$ the vector of Pauli matrices, and $V_\mathrm{dis}({\mathbf x})$ the scattering potential of the disorder.
We use in our calculations the effective mass of CdTe $m^*=0.1m_e$ where $m_e$ is the bare electron mass.
We assume that at low temperature the material has a giant Zeeman splitting\cite{Furdyna1988} with
a very large ${g}_{\mathrm{eff}}$, hence we use here ${g}_{\mathrm{eff}}$ ranging from 177 to 550.
Disorder is modeled with an Anderson-like impurity model
to account for spin-independent scattering processes.\cite{ando}

We study both finite rectangular waveguides as well as periodic systems in the direction transverse to
the transport direction ($y$-direction).
With periodic boundary conditions we emulate wide systems which would otherwise be beyond computational capabilities.
In both cases we calculate magnetoconductance in a domain of length $L$ and width $W$
and the waveguide is connected to leads at $x=0$ and $x=L$.
Besides the helical magnetic field we assume a tunable homogeneous magnetic field ${\mathbf B}_{{\mathrm{c}}}=B_{\mathrm{c}}( 0, 0,-1)$
perpendicular to the 2DEG plane. In the leads this gives rise to a significant spin polarization
\begin{equation}
p=(n_\uparrow-n_\downarrow)/(n_\uparrow+n_\downarrow),
\end{equation}
where $n_\sigma$ denotes the number of occupied modes for spin $\sigma=\{\uparrow,\downarrow\}$.

\subsection{Magnetic field texture}

The magnetic field in the calculations has a rotating component
which is helical in the transport direction,
\begin{equation}
{\bf B}_{\rm helix}({\bf x})=B_{\rm helix}(\sin 2\pi x/a, 0 ,\cos 2\pi x/a)
\label{eq:helix}
\end{equation}
where $a$ is the pitch of the helix.
The Zeeman energy of the total magnetic field
${\mathbf B}({\mathbf x})={\mathbf B}_{\rm helix}({\mathbf x})+{\mathbf B}_{\rm c}$ for parallel ($+$) and antiparallel ($-$)
spin orientations (for $g_{\mathrm{eff}}>0$) is then
\begin{equation}
E_{{\mathrm{Z}},\pm}({x}) = \pm \frac{1}{2}g_{\mathrm{eff}} \mu_B B_{\mathrm{helix}} \sqrt{1 + \gamma^2
+ 2 \gamma \cos(2\pi x/a)},
\label{eq:ez}
\end{equation}
where $\gamma = B_{\mathrm{c}}/B_{\mathrm{helix}}$.
The direction of the total field is
\begin{equation}
\theta(x)=\arctan\left(\frac{\sin (2\pi x/a)}{\cos (2\pi x/a)+\gamma}\right).
\label{eq:theta}
\end{equation}

We study spin transmission in the regime where spin transport is predominantly adiabatic.\cite{Born1928}
The magnetic field in the electron's frame of reference changes then slowly
on time scales of the order of the period of Larmor spin precession $2\pi/\omega_\mathrm{L}=2\pi \hbar/(g_{\mathrm{eff}}\mu_B B)$.
Denoting the magnetic field modulation frequency in the electron's frame of reference by $\omega_\mathrm{mod}=2\pi v_\mathrm{F}/a$
we use $Q = \omega_\mathrm{L}/\omega_\mathrm{mod}$ as a measure of the degree of adiabaticity.\cite{berry-phase,semiclassic}
In ballistic systems the adiabatic regime is $Q \gg 1$. In the presence of disorder the condition is $Q \gg a/l_e$, where
$l_e$ is the electron mean free path.\cite{Popp2003}

\subsection{Landau-Zener model}

In the absence of the homogeneous field component $\theta(x)=2\pi x/a$, and
for electrons moving parallel to the helix axis
\begin{equation}
Q_{\mathrm{helix}}=\mu_{B}g_{\mathrm{eff}}B_{\mathrm{helix}}a/(2\pi\hbar v_{\mathrm{F}})
\label{q-helix}
\end{equation}
is constant.
However, if both field components are present $\theta(x)$ changes faster close to $x=(n'+\frac{1}{2})a$, where $n'$ is an
integer (see Fig.~\ref{fig:fig0}b):
\begin{equation}
\frac{\partial\theta(x)}{\partial x}=\frac{\frac{2\pi}{a}(1+\gamma\cos (2\pi x/a))}{1+\gamma^{2}+2\gamma\cos (2\pi x/a)}.
\end{equation}
The angle $\theta(x)$ is discontinuous at $x=(n'+\frac{1}{2})a$ in the limit $\gamma\to 1$, and conditions for adiabaticity are therefore violated.\cite{betthausen}
The Zeeman-split levels are then intertwined and they cross at $x=(n'+\frac{1}{2})a$ (see Fig.~\ref{fig:fig0}c).
In this limit a spin wave function which has been transported adiabatically
to $x=(n'+\frac{1}{2})a$ has an overlap of 1 with the upper band with vanishing energy difference between the bands. This means that
a transition occurs to the upper band with probability 1.
For spin-polarized states the upper band is at least partially above the Fermi energy and the electron's
wave function decays. This leads to spin backscattering.
Spin-compensated transport modes are not affected since their energy remains below the Fermi energy.
The relative strength of the homogeneous and helical field components therefore determines
adiabaticity and the backscattering probability of spin.

The energies of spin-split eigenstates $\phi_\pm$ in a combination of homogeneous and helical magnetic fields are
given by Eq.\  (\ref{eq:ez}) and depicted for three representative values of $\gamma$ in Fig.~\ref{fig:fig0}.
They form a two-level system where diabatic transitions are possible between the states.
Landau, Zener, St\"uckelberg and Majorana
calculated the diabatic transition probability in particular  two-level systems.\cite{landau,zener-transition,stueckelberg,majorana}
The levels given by Eq.\ (\ref{eq:ez}) anticross at $x=(n'+\frac{1}{2})a$, and a diabatic transition from $\phi_-$ to $\phi_+$ occurs with a probability
\begin{equation}
P=\exp\left(-\frac{2\pi}{\hbar^{2}}\epsilon_{12}^{2}/\alpha\right),
\label{eq:landau-zener}
\end{equation}
where the non-diagonal energy term
\begin{equation}
\epsilon_{12}=\frac{1}{2}\mu_{B}g_{\mathrm{eff}}B_{\mathrm{helix}}|\gamma-1|
\label{epsilon}
\end{equation}
equals half the closest distance between the eigenenergies at the closest approach
and $\alpha=\frac{1}{\hbar}\frac{d}{dt}(\epsilon_{+}-\epsilon_{-})$
measures how fast the energies of eigenstates $\phi_{+}$ and $\phi_{-}$ approach each other
during spin transport at the level anticrossing.
This depends on the transport velocity. Assuming that an electron moves parallel to the helix axis at speed $v_{\mathrm{F}}$,
the energy difference $\epsilon_{+}-\epsilon_{-}$ can be approximated from
the eigen\-energies in the limit $\gamma \to 1$ yielding
\begin{eqnarray}
\alpha &=&\frac{1}{2}\mu_{B}g^{\mathrm{eff}}B_{\mathrm{helix}}\frac{1}{\hbar}\frac{2\pi}{a}v_{\mathrm{F}}\lim_{\tau\to\pi^-}\frac{d}{d\tau}\sqrt{2+2\cos\tau} 
\nonumber\\
&=&\mu_{B}g^{\mathrm{eff}}B_{\mathrm{helix}}\frac{1}{\hbar}\frac{2\pi}{a}v_{\mathrm{F}},\label{alpha}
\end{eqnarray}
where $\tau=\frac{2\pi}{a}v_{\mathrm{F}} t$. In waveguides electrons have a component of
momentum perpendicular to the helix axis and effectively $\alpha$ is lower.
Using Eqs.\ (\ref{eq:landau-zener}),  (\ref{epsilon}), and (\ref{alpha}) the probability of diabatic transition
can be approximated as
\begin{equation}
P\approx\exp\left(-\pi(\gamma-1)^{2}Q_{\mathrm{helix}}/2\right),
\label{l-z}
\end{equation}
where $Q_{\mathrm{helix}}$ is given by Eq.\ (\ref{q-helix}).

The transition amplitude
can also be obtained within the formalism introduced by Dykhne for time-dependent Hamiltonians.\cite{dykhne,davis-pechukas}
However, we found that the transition probability in our case of predominantly adiabatic transport does
not significantly differ from the Landau-Zener formula (\ref{eq:landau-zener}).
%bulk system: ndown, nup
%gamma=0: 109 151
%gamma=2: 46 188

\section{Results}

\subsection{Numerical method}
\label{sec:methods}

The magnetoconductance of waveguides with spin-polarized states
is calculated using a recursive Green's function (RGF) algorithm 
based on a tight-binding discretization of the system.\cite{mike} Moreover, we compare
the results to the Landau-Zener approximation for ballistic systems.
The electron mean free path $l_e$ is estimated from the disorder strength.
The transmission coefficients $t_{nm}$ of transport modes are calculated with the RGF algorithm
and conductance $G$ is obtained from the Landauer formula
\begin{equation}
G(B)  = G^0 \sum_{n,m,\sigma, \sigma^{\prime}} \left|t^{\sigma \sigma^{\prime}}_{nm}(B)\right|^2 \, ,
\label{landauer}
\end{equation}
where $G^0=e^2/h$ is the conductance (per spin) of one channel,
$\sigma$ and $\sigma'$ denote the spin indices, and $n$ and $m$ the channel indices.
In both leads there is a homogeneous magnetic field $B_{\mathrm{c}}+B_{\mathrm{helix}}$ perpendicular
to the 2DEG surface.
In disordered systems $t_{nm}$ is averaged over random disorder configurations in our calculations.
The number of configurations ranges from 10 configurations in large bulk-like multi-mode systems
to more than 100\,000 in single mode waveguides where the electron mean free path is short.

\subsection{Tuning of spin backscattering with Landau-Zener transitions}
\label{sec:tuning}

\begin{figure*}
\includegraphics[width=1.75\columnwidth]{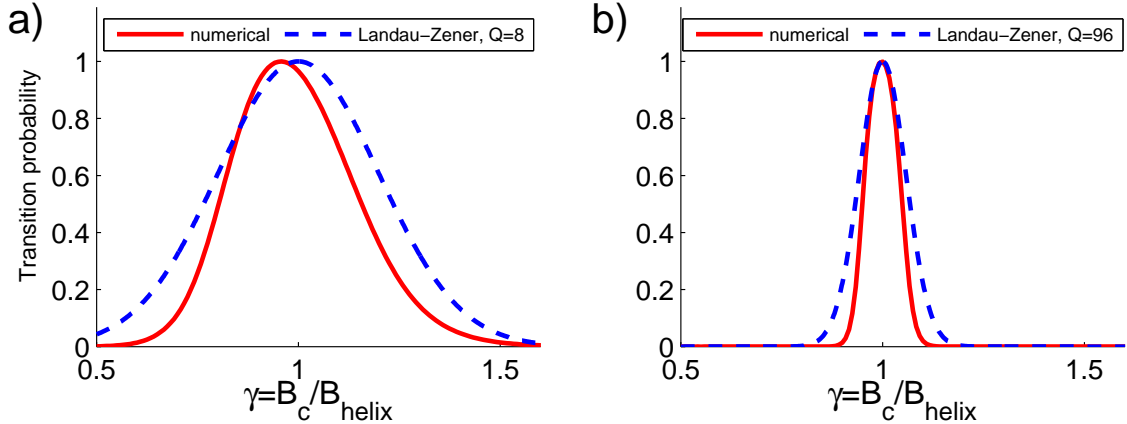}
\caption{(Color online) Diabatic transition probability in a ballistic single-mode waveguide of one helical pitch $L=a$
calculated with the RGF algorithm (solid lines) at $a=1.5\;\mathrm{\mu m}$ corresponding to $Q=8$ (a) and
$a=18\;\mathrm{\mu m}$ corresponding to $Q=96$ (b).
In the latter case the helical magnetic field changes more slowly in the electron's frame of reference and transport is more adiabatic.
Results are compared to the transition probability Eq.\ (\ref{l-z}) from the Landau-Zener approximation (dashed lines). The adiabaticity
parameter $Q$ is approximated using Eq.\ (\ref{q-helix}).
}
\label{fig:fig1}
\end{figure*}

\begin{figure*}
\includegraphics[width=1.75\columnwidth]{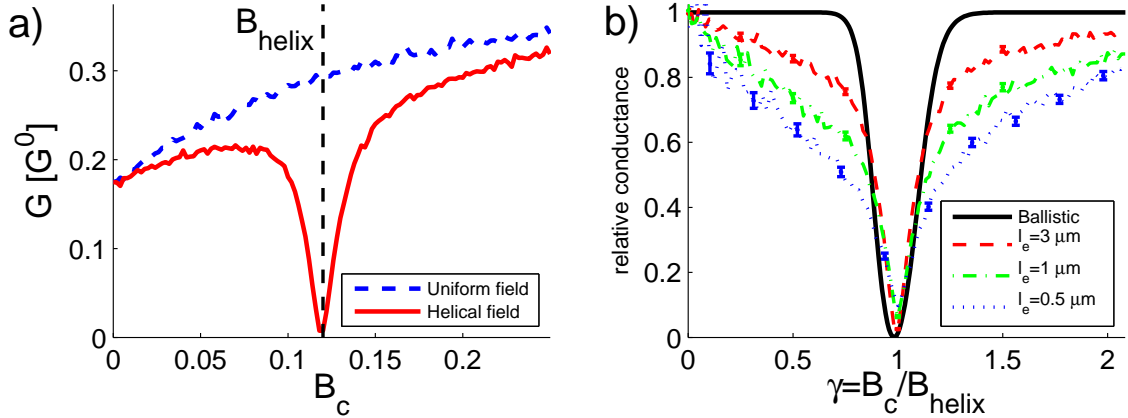}
\caption{(Color online)
Calculated disorder-averaged transmission in a spin-polarized single-mode waveguide
with a helical field of strength $B_{\mathrm{helix}}=0.12\;\mathrm{T}$
and pitch $a=L=1\;\mathrm{\mu m}$.  
Waveguide width is $24\;\mathrm{nm}$, $m^*=0.1m_e$, $E_\mathrm{F}=6.6\;\mathrm{meV}$, and $g_{\mathrm{eff}}=177$.
a) Magnetoconductance shows a dip associated with spin
backscattering at $B_{\mathrm{c}}=B_{\mathrm{helix}}$.
Magnetoconductance in a homogeneous field is shown for comparison.
Electron mean free path $l_e=3\;\rm{\mu m}$.
b) Relative magnetoconductance for various electron mean free paths ($l_e=0.5, 1, 3\;\mathrm{\mu m}$).
Conductance is normalized to conductance in a homogeneous magnetic
field of the same strength. The energy levels for the ballistic system are shown in
Fig.\ \ref{fig:fig0}.
}
\label{fig:fig2}
\end{figure*}

\begin{figure}
\includegraphics[width=0.9\columnwidth]{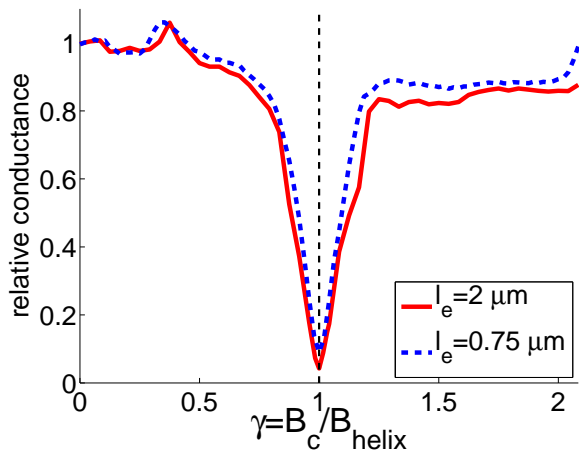}
\caption{(Color online) Relative magnetoconductance in a multi-mode disordered waveguides with 8 spin-polarized transport modes at $\gamma=1$.
Conductances are normalized to respective conductances in a homogeneous magnetic field of the same strength.
Waveguide length is one helical pitch $L=a=1\;\mathrm{\mu m}$ and $E_{\mathrm{F}}=8\;{\mathrm{meV}}$.
Electron mean free path $l_e$ is indicated in the figure.
}
\label{fig:fig3}
\end{figure}

\begin{figure}
\includegraphics[width=0.9\columnwidth]{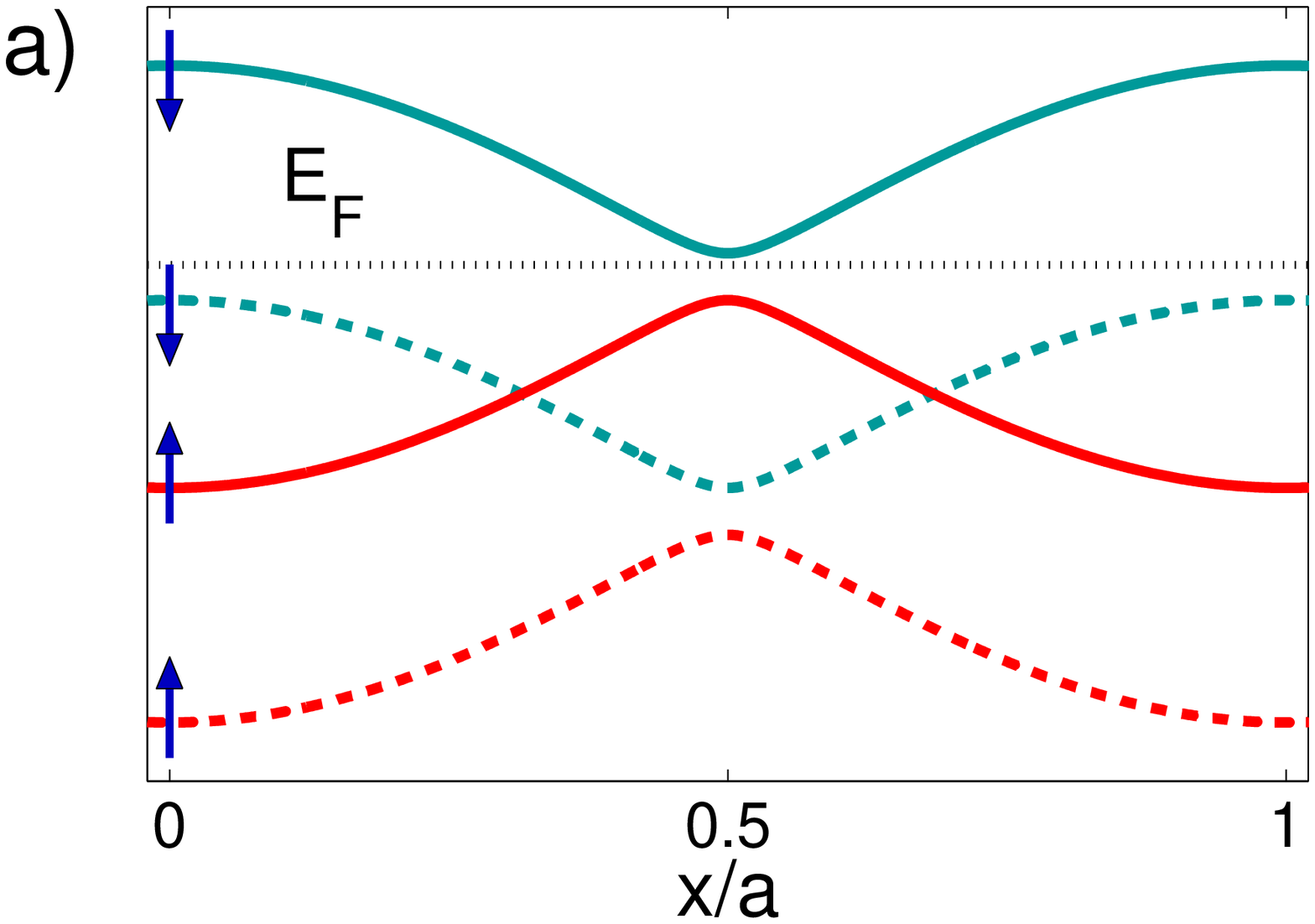}
\includegraphics[width=0.9\columnwidth]{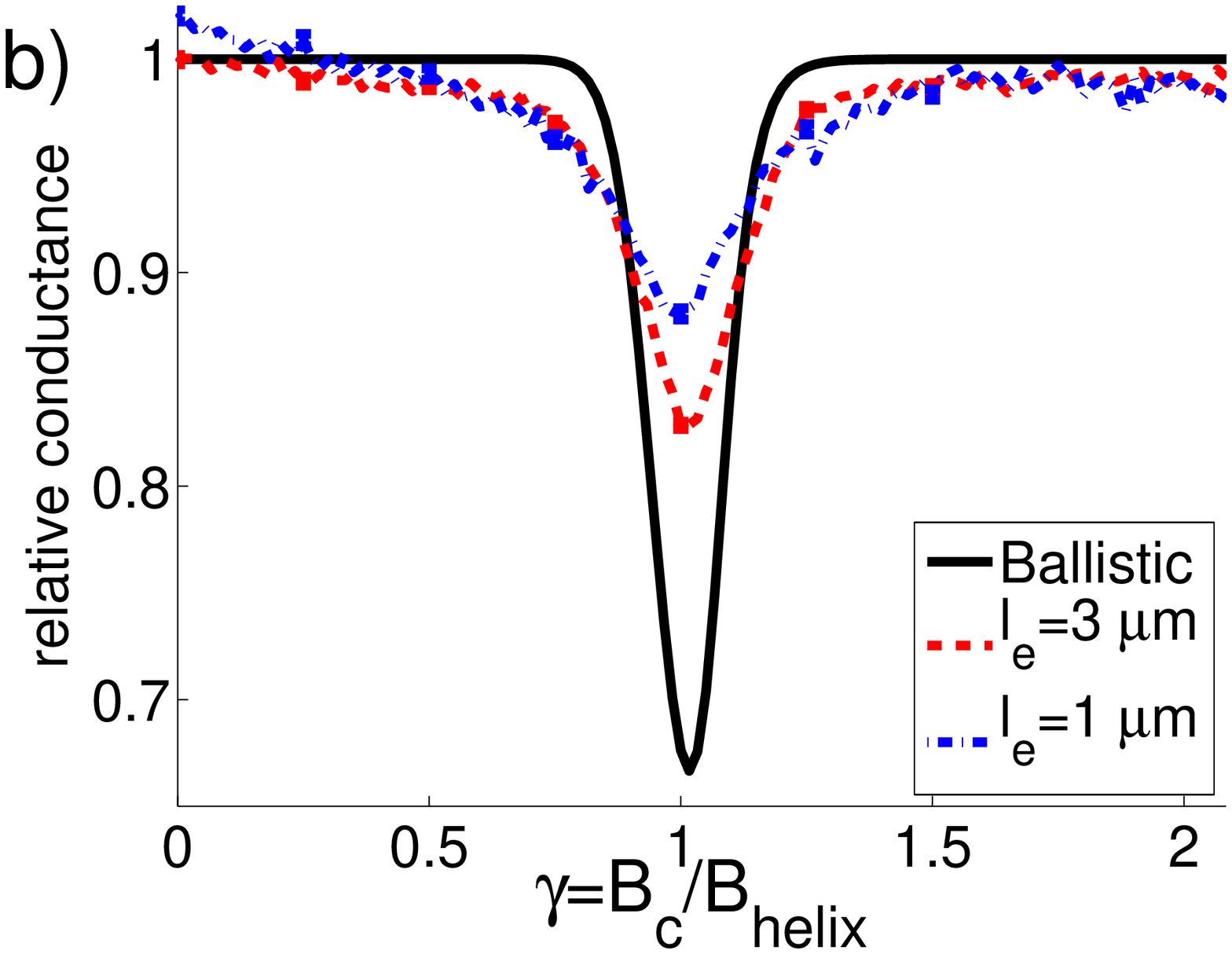}
\caption{(Color online) a) In a multi-mode waveguide energies of the Zeeman-split eigenstate pairs are either both below the Fermi energy (dashed lines)
giving rise to a spin-compensated mode, or the higher Zeeman-split eigenstate (solid lines) is above the Fermi energy $E_\mathrm{F}$
giving rise to a spin-polarized mode.
Eigenstates are plotted at $\gamma=0.8$. The arrows show the corresponding spin directions at $x=0$.
b) Relative magnetoconductance  calculated in a double-mode ballistic waveguide (solid line) and disordered waveguides (dashed line for
$l_e=3\;\mathrm{\mu m}$ and dash-dotted line for $l_e=1\;\mathrm{\mu m}$).
The energy levels of the modes are depicted in a); the lower mode is spin compensated and the upper mode is spin polarized, $p=0.33$.
The system parameters are otherwise the same as in the caption of Fig. \ref{fig:fig3}.
}
\label{fig:figmultibans}
\end{figure}

\begin{figure}
\includegraphics[width=1\columnwidth]{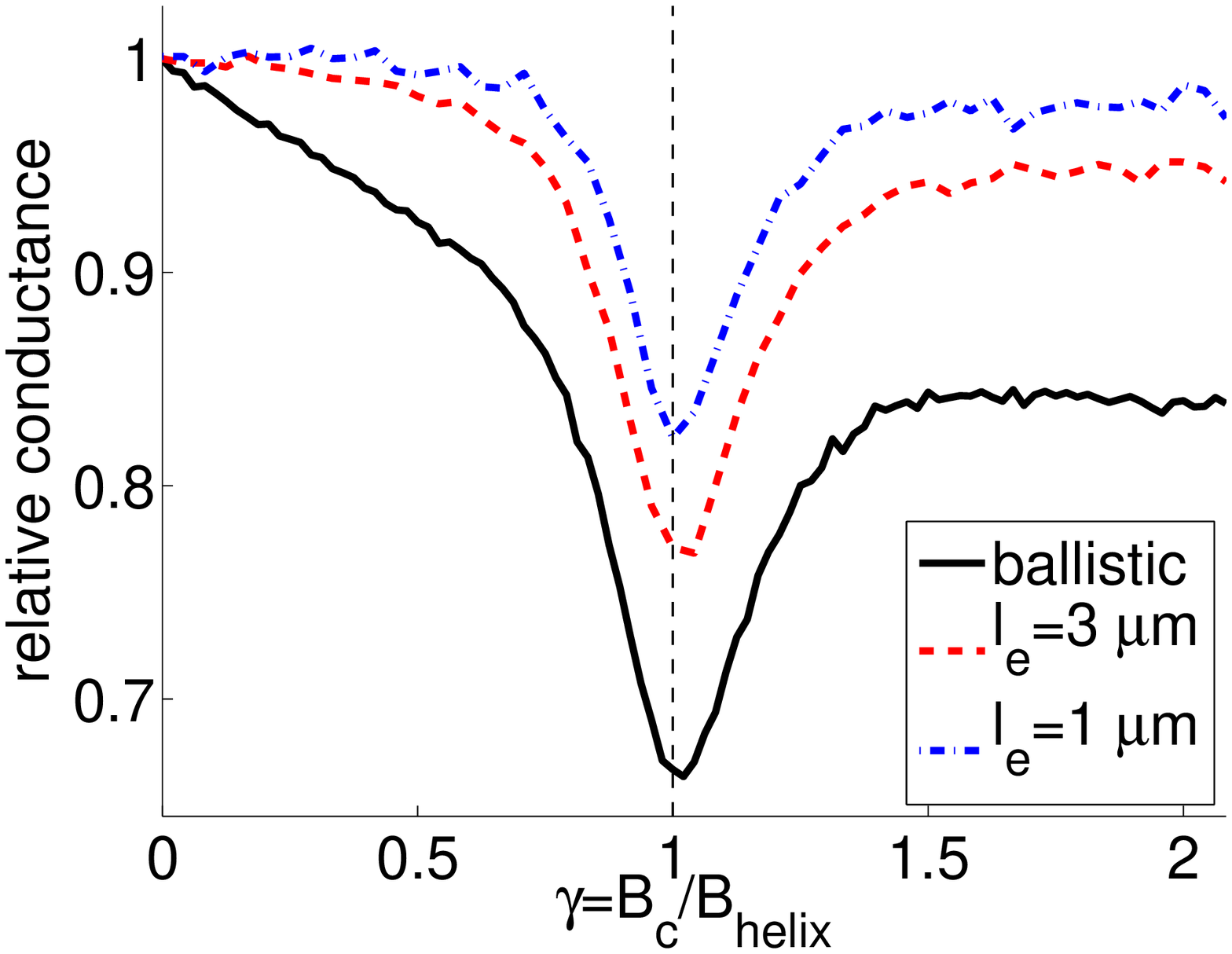}
\caption{(Color online) Relative magnetoconductance in a partially polarized ($p=0.34$ at $\gamma=1$) 
disordered waveguide with 170 transverse modes. Waveguide length $L=a=1\;\mathrm{\mu m}$  and width is $3\;\mathrm{\mu m}$.
Conductance is normalized to conductance in a homogeneous magnetic field of the same strength.
$E_{\mathrm{F}}=6.6\;{\mathrm{meV}}$ and $g_{\mathrm{eff}}=550$.
}
\label{fig:fig4}
\end{figure}

We study first transitions caused by a single level (anti)crossing in the Zeeman-split bands. Waveguide length is therefore one helical pitch, $L=a$.
We omit therefore orbital effects as a first approximation and
the effective mass Hamiltonian includes only the kinetic term, the Zeeman coupling and the disorder potential,
\begin{equation}
\mathrm{\hat{H}} =
\frac{1}{2 m^*}\ \hat{\bf{P}}^2 + \frac{1}{2} {g}_{\mathrm{eff}} \mathrm{\mu_{\mathrm{B}}} \mathbf{B}(\mathbf{x})\cdot \bm{\sigma}+V_\mathrm{dis}({\mathbf x}).
\label{eq:spinonly}
\end{equation}
In long waveguides with sequences of level (anti)crossings orbital effects become important and they are studied in Sec. \ref{sec:periodic}.
In the case of rectangular waveguides we assume
an infinite potential well of width $W$
in the transverse direction. For mode $n$ the quantum well energy is 
$E_n = ({\hbar^2}/{2 m^* })({\pi^2 n^2}/{W^2})$ for $n = 1, 2, 3, \ldots$
In the simplest case transport involves only one spin-polarized mode ($n=1$).
The spin-splitting of this mode, Eq.\ (\ref{eq:ez}), in the modulated
magnetic field gives rise to a two-level system with a periodic sequence of
level (anti)crossings (see Fig.\ \ref{fig:fig0} for one period).

In ballistic systems the transition probability from the lower spin eigenstate to the higher one
can be calculated either using the Landau-Zener approximation Eq. (\ref{l-z}) or the RGF algorithm. 
For spin-polarized states the upper band is at least partially above the Fermi energy and the wave function
decays after a diabatic transition. 
This leads to spin backscattering.
Figure \ref{fig:fig1} shows the probability of spin backscattering
in a single-mode wave function calculated with both methods.
In the low $Q$ regime numerical results show a shift in the peak position from $\gamma=1$ towards lower $\gamma$ values (Fig.\ \ref{fig:fig1}a).
Spin transport is not perfectly adiabatic in this regime and spin is slightly non-aligned with the magnetic field resulting in precession.
At $\gamma=1$ in the $Q\gg 1$ regime the probability of diabatic transition and spin backreflection tends to 1 (Fig.\ \ref{fig:fig0}c).
The adiabatic theorem is reflected in the probability distribution which gets narrower with increasing $Q$.

The role of spin-independent disorder scattering was analyzed within the RGF formalism. Figure \ref{fig:fig2}a shows
a dip in the magnetoconductance associated with spin transmission blocking in a disordered single-mode wave\-guide.
We normalize magnetoconductances in the following figures
to the corresponding values in a homogeneous magnetic field of strength $B_{\mathrm{c}}+B_{\mathrm{helix}}$
%$G_{\mathrm{rel}}(B)=G_{\mathrm{helix}}(B_{\mathrm{c}}+B_{\mathrm{helix}}/G_(B_{\mathrm{c}}+B_{\mathrm{helix}})$
in order to factor out ohmic resistance caused by the disorder.
The magnetoconductance calculations at different mean free paths show that almost all transmission
is blocked at $B_{\mathrm{c}}=B_{\mathrm{helix}}$ even if the mean
free path is shorter than the magnetic field helix pitch $a$
(Fig.\ \ref{fig:fig2}b).
The result can be understood in terms of adiabatic spin transport which
keeps spin aligned with the magnetic field despite scattering from disorder. 
At $\gamma=1$ there is no adiabatic path through the system and spin is reflected (Fig.\ \ref{fig:fig0}c).
With increasing disorder the dip in the relative conductance broadens.
Since electrons scatter from impurities they pass the level (anti)crossing many times which enhances backreflection probability.

The result applies also to multiple spin-polarized channels.
Figure \ref{fig:fig3}
shows that in a spin-polarized multi-channel system current is almost completely
switched off at $\gamma=1$
even in the presence of disorder. There is a small leakage current through the system at $\gamma=1$
because conditions of adiabaticity hold only approximately ($Q\approx 11$ at $n=1$ and $Q\approx 25$ at $n=8$ in this case)
and spin flips are therefore possible.
%EZ=g muB*B=1000*bohr magneton*0.12 T=5.788E-5*1000*0.12 eV
%W=140 nm
%mode=1 or 8
%EF=6.6
%g=1000
%B=0.12 T
%a=1000 nm
%hbar^2/(2*electron mass*0.1)*(pi^2*8^2)/(140^2*nanometers^2)/electron volt
%Ekintrans=0.19 or 6.9
%Ekin=6.41 meV or 1.2 meV
%vF=sqrt(2*6.41/1000*electron volt/(0.1*electron mass))=150160
%1000*0.12*0.0000578*electron volt/hbar*1000*nanometers/(2*pi*(150160*meter/second))=11
%vF=sqrt(2*1.2/1000*electron volt/(0.1*electron mass))=65000
%1000*0.12*0.0000578*electron volt/hbar*1000*nanometers/(2*pi*(65000*meter/second))=25
The mean free paths $l_e$ and the ratios $l_e/a$ in the calculations are of the order of those which are attained in
(Cd,Mn)Te quantum wells.\cite{betthausen}

In partially spin-polarized systems
Zeeman-split eigenstates $\phi_\pm$ of spin-compensated modes both remain below the Fermi energy (see dashed lines in Fig.\ \ref{fig:figmultibans}a).
Spin backscattering at a level (anti)crossing
does therefore not occur and both spins are transmitted in the ballistic case. This is shown for a double-mode ballistic calculation in Fig.\ \ref{fig:figmultibans}b
where the upper spin-polarized mode is reflected at $B_{\mathrm{c}}=B_{\mathrm{helix}}$ but the lower spin-compensated mode
gives $2G^0$ conductance (the normalized relative conductance is therefore 2/3).
Note that if $E_{\mathrm{F}}$ is below the maximum value of the spin-polarized energy band (solid red line in Fig.\ \ref{fig:figmultibans}a) the wave function of these
states decay without a Landau-Zener transition.

In disordered waveguides with partial
spin polarization ($p<1$) the resistance is increased partly due to spin backscattering at Landau-Zener transitions 
and partly due to disorder scattering, which affects both spin eigenstates.
The latter gets more important as disorder increases and there are also transitions between spin-polarized and
spin-compensated transport modes.
An electron which is initially in a spin-polarized mode may then scatter
to another mode which is not backscattered at a level (anti)crossing
(e.g. the mode shown with dashed lines in Fig.\ \ref{fig:figmultibans}a). 
As a consequence the electron may transmit and the relative conductance dip at $\gamma=1$ decreases.
The total relative conductance depends on the disorder strength as shown in Fig.\ \ref{fig:figmultibans}b.

The above results are directly applicable to bulk-like multi-mode systems. Figure\ \ref{fig:fig4} shows magnetoconductance
in a partially polarized multi-mode system ($n=170$) where a similar conductance pattern develops due to spin backscattering.
The magnetoconductance is asymmetric with respect to $\gamma=1$ since the calculations are done at constant $B_\mathrm{helix}$
and therefore spin polarization of the leads increases with $\gamma$.

Although electron may scatter at impurities, the spin still
aligns with the local external magnetic field if it changes slowly in the electron's rest frame ($Q\gg 1$).
In the diabatic transport regime ($Q \ll 1$) the spin wavefunction becomes a superposition of local eigenstates which leads to spin precession in the
local magnetic field. The above described way to control spin transmission is then not possible.

\subsection{Sequences of level (anti)crossings}
\label{sec:periodic}

\begin{figure}
\includegraphics[width=1.\columnwidth]{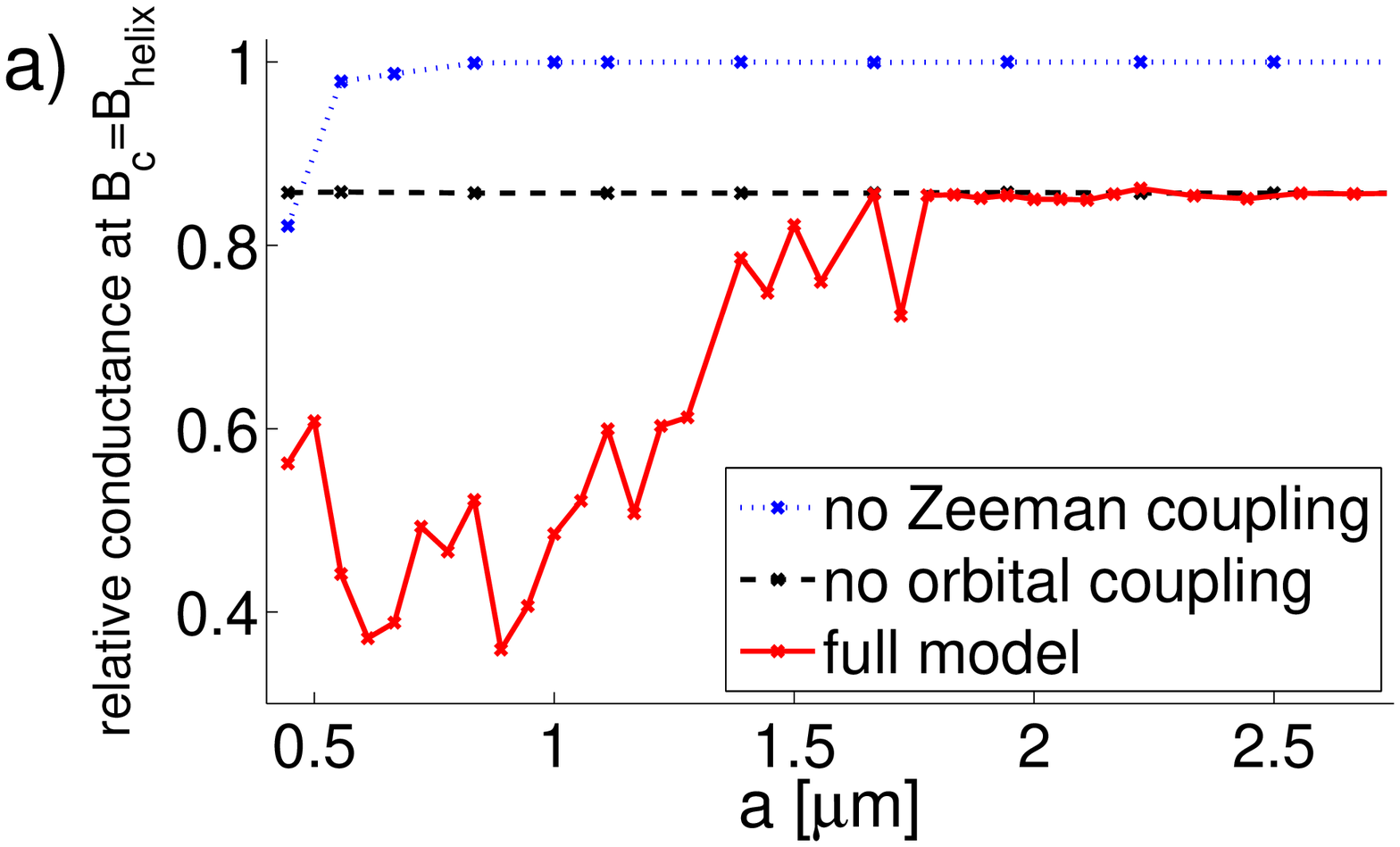}
\includegraphics[width=1.\columnwidth]{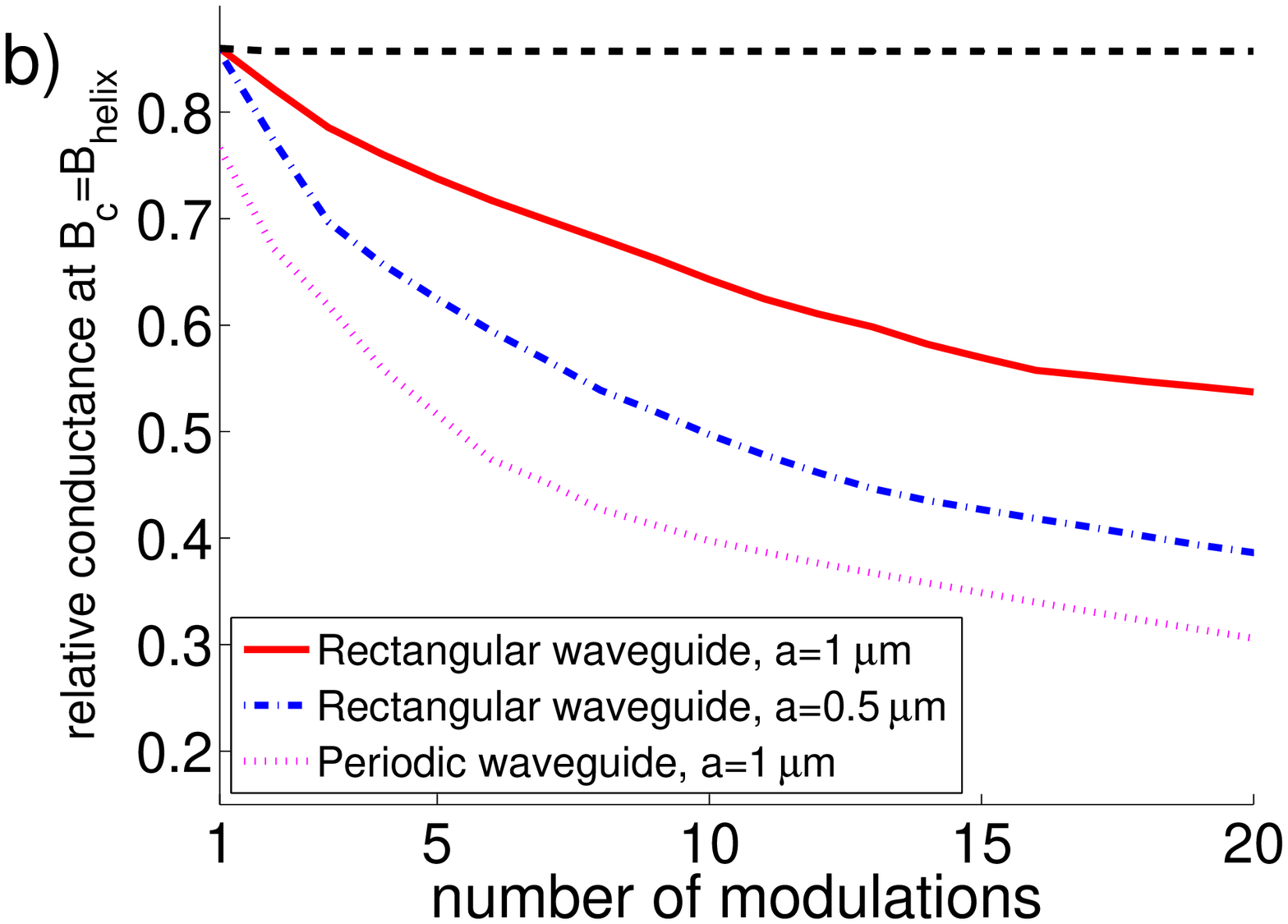}
\caption{(Color online) Relative magnetoconductance in long ballistic waveguides with $n'$ helical modulations.
Magnetoconductance is calculated at $B_{\mathrm{c}}=B_{\mathrm{helix}}=0.12\;\mathrm{T}$, $g_{\mathrm{eff}}=177$, and
$p=0.14$. 
a) Relative magnetoconductance as a function of the helix pitch $a$ in the case $n'=16$.
Figure shows relative magnetoconductance calculated with the full model Hamiltonian (Eq. (\ref{eq:h}), solid line),
in the absence of the Zeeman coupling ($g_{\mathrm{eff}}=0$, dotted line), and in the absence of the orbital coupling (${\bf A}={\bf 0}$, dashed line).
Adiabaticity parameter $Q\approx 1$ for the mode $n=1$ at $a=0.5\;\mathrm{\mu m}$.
Waveguide width is $W=425 \;\mathrm{nm}$, number of transport modes $n=16$, $E_{\mathrm{F}}=7.4\;{\mathrm{meV}}$.
b) Relative magnetoconductance as a function of $n'$ in a rectangular waveguide
(for $a=1\;\mathrm{\mu m}$, solid line, and for $a=0.5\;\mathrm{\mu m}$, dash-dotted line)
and in a periodic systems in the $y$-direction for $a=1\;\mathrm{\mu m}$.
Full model Hamiltonian is used here. The dashed line shows the relative magnetoconductance in the adiabatic limit.
Waveguide width is $W=350 \;\mathrm{nm}$.}
\label{fig:fig4.5}
\end{figure}

%\begin{figure}
%\caption{(Color online) Relative magnetoconductance in a partially spin polarized ($p=0.14$) ballistic waveguide at $\gamma=1$
%as a function of number $n'$ of magnetic modulations, i.e. number of helix pitches $a$,
%for rectangular geometries and periodic systems in $y$-direction. The orbital coupling is included in the calculations.
%Adiabaticity parameter $Q\approx 1$ for the mode $n=1$ at $a=0.5\;\mathrm{\mu m}$.
%The dashed line shows the relative magnetoconductance in the adiabatic limit.
%Waveguide width $W=350\;\mathrm{nm}$, $g_{\mathrm{eff}}=177$, $E_{\mathrm{F}}=6.6\;{\mathrm{meV}}$, $B_{\mathrm{helix}}=0.12\;{\mathrm{T}}$
%and the number of transport modes is 16.
%g=177
%a=500 nm
%W=350 nm
%mode=1
%EF=6.6
%B=0.12 T
%Ekintrans=0.19
%Ekin=6.41 meV
%vF=sqrt(2*6.41/1000*electron volt/(0.1*electron mass))=150160
%177*0.12*0.0000578*electron volt/hbar*500*nanometers/(2*pi*(150160*meter/second)) ~ 1
%}
%\label{fig:fig5}
%\end{figure}
\begin{figure}
\includegraphics[width=0.9\columnwidth]{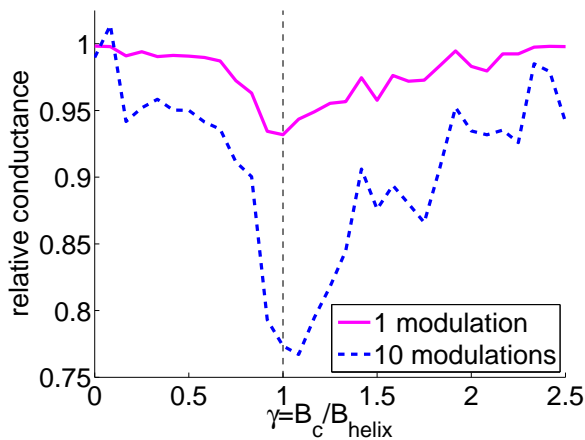}
\caption{(Color online) Relative magnetoconductance in a partially
spin-polarized ($p=0.11$) disordered waveguide for one helical modulation
(solid line) and for 10 modulations (dashed line). Magnetic field helix
pitch is $a=0.5 \;\mathrm{\mu m}$ in both cases and $Q\approx 1$ for the mode $n=1$.
Electron mean free path $l_e=3\;\mathrm{\mu m}$,
waveguide width $W=350\;\mathrm{nm}$, $g_{\mathrm{eff}}=177$, $E_{\mathrm{F}}=6.6\;{\mathrm{meV}}$, $B_{\mathrm{helix}}=0.12\;{\mathrm{T}}$
and the number of transport modes is 16.
}
\label{fig:fig5b}
\end{figure}
\begin{figure}
\includegraphics[width=1.0\columnwidth]{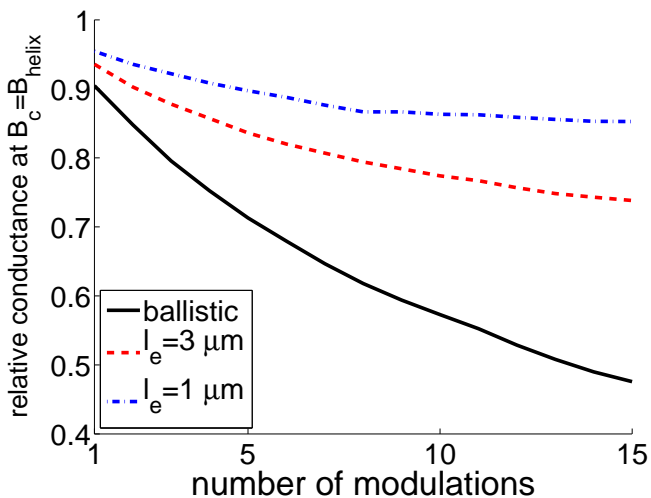}
\caption{(Color online) Relative magnetoconductance in a partially spin-polarized
($p=0.11$) disordered waveguide at $\gamma=1$
as a function of number of magnetic modulations (in helix pitches
$a=0.5 \;\mathrm{\mu m}$). 
The waveguide parameters are otherwise the same as in Fig.\ \ref{fig:fig4.5}b.
}
\label{fig:fig6}
\end{figure}
A single (anti)crossing in the Zeeman-split energy levels has a spin transmission blocking effect
as shown in Sec. \ref{sec:tuning}. This causes an increase in resistance which depends on spin polarization and disorder strength.
Resistance modulation is enhanced if electrons are transported through a sequence of level (anti)crossings ($n'$ helical modulations, $L=n'a$).
The electron transmission probability depends
on transitions between transport channels caused by disorder scattering or orbital dynamics in the magnetic field.
Hence we take orbital effects into account and use Hamiltonian (\ref{eq:h}) to calculate magnetoconductance with the RGF method.

In ballistic systems there are no transitions between transport channels in the adiabatic transport limit since
the local transverse modes change slowly with the magnetic field.
However, if the local magnetic field in the electron's frame of reference 
changes rapidly, transitions between the modes may lead to
reoccupation of a backscattered spin-polarized mode
(e.g. in Fig.\ \ref{fig:figmultibans}a these transitions would be from the spin-compensated
modes (dashed lines) to spin-polarized modes (solid lines)).
The electron may subsequently backscatter in the following level (anti)crossing. The relative magnetoconductance therefore
decreases with magnetic field helix pitch at $\gamma=1$ (see Fig.\ \ref{fig:fig4.5}a).
Neither a pure Zeeman coupling nor orbital coupling alone account for the clearcut reduction in the relative
conductance for $a\le 1.5\;{\mathrm{\mu m}}$.
For magnetoconductance traces as a function of magnetic field in the ballistic case see supplementary material
in Ref.\ \onlinecite{betthausen}.

Figure \ref{fig:fig4.5}b shows magnetoconductance in a partially polarized ballistic waveguide as a function of the number of helical modulations $n'$
in the Zeeman-split energy bands. Conductances are calculated at 
$\gamma=1$ where the diabatic transition probability is highest.
The degree of adiabaticity is lower in the short helix pitch $a=0.5\;\mathrm{\mu m}$ and
the probability of mode transitions is higher. The relative conductance decrease is amplified with increasing $n'$
and results in a huge dip in magnetoconductance at $\gamma=1$ if the number of modulations is large.
We find qualitatively similar but quantitatively larger effects in periodic systems (dotted line in Fig.\ \ref{fig:fig4.5}b).

The above mechanism causes enhanced
spin blocking also in disordered waveguides.
The resistance is effectively higher for the spin-polarized channels than for spin-compensated channels due
to diabatic transitions and spin backscattering.
The dip in the relative magnetoconductance at $\gamma=1$ increases with the number of magnetic modulations (Fig.\ \ref{fig:fig5b}).
This is in line with the experiments in Ref.\ \onlinecite{betthausen}.
Figure \ref{fig:fig6} shows the relative conductance at $\gamma=1$ in disordered waveguides with spin polarization $p=0.11$.
We note that the relative conductance change at $n'=15$ for $l_e={1\;\mathrm{\mu m}}$ is larger
than the relative conductance change at $n'=1$ in the ballistic case.

\section{Conclusions and outlook}
\label{sec:conclusions}

Our results show that spin transistor action can be realized via tunable Landau-Zener transitions.
The mechanism is tolerant against spin-independent disorder scattering
for an Anderson impurity model. Completely spin-polarized systems
show full spin backscattering, and thus current switching,
even when the mean free path of electrons is of the order of the magnetic modulation length.

In partially spin-polarized waveguides the resistance modulation decreases with
increasing disorder strength. 
However, the resistance modulation due to Landau-Zener transitions can be enhanced with a sequence
of (anti)crossings in the spin-split bands. Orbital transitions
cause successive reoccupation and backscattering of spin-polarized modes.
This effect provides also an explanation why the spin blocking
effect in experiments is larger than the theoretical prediction for ballistic systems
in the absence of  orbital effects.\cite{betthausen}

Implementation of  a spin transistor mechanism via tunable Landau-Zener transitions might be a more
feasible approach to realize spin transistor functionality than
controlling spin dephasing times using an interplay of Rashba and Dresselhaus spin-orbit couplings.\cite{loss}
In the latter proposal the transistor operation is based on the persistent spin helix state\cite{bernevig}
which is also tolerant against spin-independent disorder scattering.
However, device operation requires a delicate adjustment of the spin-orbit parameters.
Moreover, the spin-splitting is bounded by the Dresselhaus spin-orbit coupling strength that 
depends on the crystal lattice structure.

Several technical challenges remain before our concept can be realized in a useful spin transistor
device. The magnetic fields for spin transmission control could be generated with magnetic gates
(see supplementary material in Ref.\ \onlinecite{betthausen}).
The giant Zeeman effect in known materials is significant only at low temperatures.
Nevertheless, the presented spin-blocking mechanism can be applied also for other spin-splitting
interactions which persist to higher temperatures. For a more thorough discussion
of the device development aspects we refer to Ref. \onlinecite{betthausen}.
Our concepts may also be applied to other materials where helical spin ordering is present, such
as the interface of multiferroic oxides.\cite{berakdar}
To conclude, robustness of the spin blocking effect via tunable Landau-Zener transitions provides a promising alternative
strategy to design spin transistor functionality with enhanced efficiency and disorder tolerance.

\begin{acknowledgments}
We thank M.~Wimmer for providing the
code for the recursive Green's function transport equation solver, V.~Krueckl for help with implementing the periodic boundary conditions,
C.~Betthausen for careful reading of the manuscript and comments, and D.~Weiss for many helpful discussions.
We acknowledge financial support from the Deutsche Forschungsgemeinschaft
through SFB 689 (H. S., K. R.) and Elitenetzwerk Bayern (T. D.). 
\end{acknowledgments}

\end{document}